\documentclass[doublecol]{epl2} % for 2 columns style with line numbers
\usepackage{dsfont,amsmath,ulem,epsfig,stmaryrd}

\makeatletter
\newsavebox{\@brx}
\newcommand{\llangle}[1][]{\savebox{\@brx}{\(\m@th{#1\langle}\)}%
  \mathopen{\copy\@brx\mkern2mu\kern-0.9\wd\@brx\usebox{\@brx}}}
\newcommand{\rrangle}[1][]{\savebox{\@brx}{\(\m@th{#1\rangle}\)}%
  \mathclose{\copy\@brx\mkern2mu\kern-0.9\wd\@brx\usebox{\@brx}}}
\makeatother

\renewcommand{\Re}{{\rm Re}}
\renewcommand{\Im}{{\rm Im}}
\newcommand{\ri}{{\rm i}}
\newcommand{\rd}{{\rm d}}
\newcommand{\re}{{\rm e}}
\newcommand{\kb}{k_{\rm B}}

\title{Green-Kubo relation for thermal radiation in non-reciprocal systems}
\shorttitle{Green-Kubo relation in non-reciprocal systems} %Insert here a short version of the title if it exceeds 70 characters

\author{F. Herz \and S.-A. Biehs}
\shortauthor{F. Herz \and S.-A. Biehs}

\institute{                    
  \inst{} Institut f\"{u}r Physik, Carl von Ossietzky Universit\"{a}t, D-26111 Oldenburg, Germany\\
}

\abstract{We rederive the Green-Kubo relation establishing a connection between the near- and far-field heat transfer between two objects out of equilibrium to the equilibrium fluctuations of these objects in an arbitrary environment. Employing the scattering approach in combination with the fluctuation-dissipation theorem, we generalize the previously derived Green-Kubo expression to the case of non-reciprocal objects and non-reciprocal environments.}

\begin{document}

\maketitle

\section{Introduction}

In the last decade the theory of fluctuational electrodynamics has been succesfully applied to describe nanoscale heat fluxes in many-body systems~\cite{PBAEtAl2011,MessinaEtAl2013,Edalatpour,Zhu2016,DongEtAl2017,EkerothEtAl2017,MuellerEtAl2017,Nikbakht2017,LatellaEtAl2017,KathmannEtAl2018,ZhuEtAl2018}. It could be shown that the heat flux between two objects can be enhanced by the presence of a third object placed in between the considered objects~\cite{PBAEtAl2011,MessinaEtAl2014,LatellaEtAl2015}. Furthermore, the heat flux can be actively tuned by adding an intermediate object~\cite{PBASAB2014,OrdonezEtAl2016,Nikbakht,Incardone2014,Nikbakht2015,GhanekarEtAl2018,KanEtAl2019}, and the heat flux can also be enhanced by coupling to the surface modes of an intermediate object or a nearby environment~\cite{Saaskilathi2014,Asheichyk2017,DongEtAl2018,MessinaEtAl2018}. In non-reciprocal many-body systems new interesting effects have been highlighted like a persistent heat flux~\cite{Zhu2016,ZhuEtAl2018,Silveirinha}, a giant magneto-resistance~\cite{Ivan2018giant,Ekeroth}, a Hall effect for thermal radiation~\cite{PBAHall,Ottreview}, and non-trivial angular and spin momentum of thermal light~\cite{OttSpin}. It could also be shown that the coupling to non-reciprocal surface modes even allows for an efficient heat flux rectification~\cite{OttetAl2019}. 

\begin{figure}
  \begin{center}
    \includegraphics[width = 0.4\textwidth]{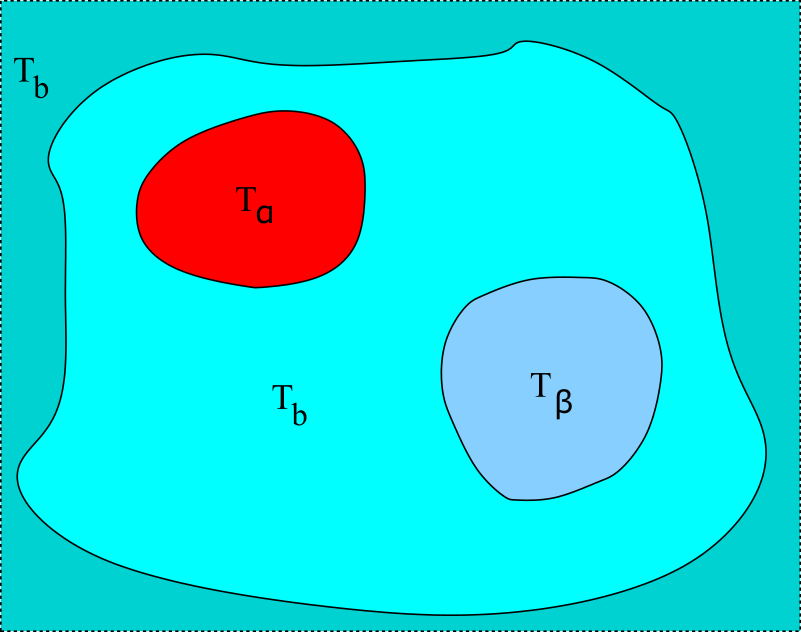}
  \end{center}
  \caption{Sketch of the considered configuration. Two compact scatterers or objects are coupled to independent heat baths which keep the objects in local thermal equilibrium at tempertures $T_{\alpha/\beta}$. Both objects are immersed in an arbitrary environment in local equilibrium with temperature $T_{\rm b}$. Here the environment is sketched as a cavity so that the cavity walls bring the environmental field into local equilibrium with temperature $T_{\rm b}$. \label{Fig:Conf}}
\end{figure}

In such many body systems, it is still possible to establish a link between the heat flux between two objects surrounded by an arbitrary environment which can consist of many other objects and the equilibrium fluctuations in the many-body system as shown by Golyk {\itshape et al.}~\cite{GolykEtAl2013} within the scattering approach. This link is given by the Kubo or Green-Kubo relation~\cite{GolykEtAl2013}. To be more precise, let us consider two compact objects of arbitrary material and shape placed within an arbitrary environment as sketched in Fig.~\ref{Fig:Conf}. When  $H^{\alpha/\beta}$ is the total power absorbed by the two objects labeled as $\alpha$ and $\beta$ having in general two different temperatures $T_{\alpha/\beta}$ 
placed in an environment which has another temperature $T_{\rm b}$, then the mean power $\llangle H^{\alpha/\beta} \rrangle$ received by 
object $\alpha$ or $\beta$ is in general finite. In global equilibrium, i.e.\ for $T_\alpha = T_\beta = T_{\rm b} \equiv T$,  $\llangle H^{\alpha/\beta} \rrangle_{\rm eq}$ must vanish. But the fluctuations do not vanish in global equilibrium in general. As shown by Golyk {\itshape et al.}~\cite{GolykEtAl2013} 
assuming that the thermal radiation fields have the Gauss property, the fluctuations in global equilibrium can be linked to the heat flux 
out of equilibrium. This link is established by the Green-Kubo relation 
\begin{equation}
   - \frac{\rd \llangle H^{\beta} \rrangle}{\rd T_\alpha} \biggr|_{T_\alpha = T_\beta = T} \!\! \!\!= \frac{1}{2 \kb T^2}\!\!\int_{-\infty}^{+\infty} \!\!\!\rd t\, \llangle[\big] H^\alpha(t) H^\beta(0) \rrangle[\big]_{\rm eq},
\label{Eq:GreenKuboGolyk}
\end{equation}
where $\kb$ is the Boltzmann constant. Note, that this relation has been derived under the assumption of ``micro-reversibility'' of the objects and the environment which means that both objects and environment are reciprocal. In this case one can use the Kubo-symmetry relation~\cite{KUBO} $\llangle[\big] H^\alpha(t) H^\beta(0) \rrangle[\big]_{\rm eq} = \llangle[\big] H^\alpha(-t) H^\beta(0) \rrangle[\big]_{\rm eq}$ to further simplify the Green-Kubo relation to 
\begin{equation}
   - \frac{\rd \llangle H^{\beta} \rrangle}{\rd T_\alpha} \biggr|_{T_\alpha = T_\beta = T} \!\! \!\!= \frac{1}{\kb T^2}\!\!\int_{0}^{+\infty} \!\!\!\rd t\, \llangle[\big] H^\alpha(t) H^\beta(0) \rrangle[\big]_{\rm eq}.
\label{Eq:GreenKuboGolyk2}
\end{equation}
This is exactly the form given in Ref.~\cite{GolykEtAl2013}.

In this letter, we will generalize this Green-Kubo relation to include the non-reciprocal case by starting the derivation without making any assumption on the ``micro-reversibility'' or reciprocity of the objects and environment. Therefore, our generalized Green-Kubo relation applies to all possible cases including the case of non-reciprocal objects and reciprocal environment, reciprocal objects and non-reciprocal environment etc. We show that our relation reproduces Eq.~(\ref{Eq:GreenKuboGolyk}) only if the two objects are reciprocal and if further the environment is reciprocal.

\section{Scattering approach}

In the following we will make use of the scattering approach together with a basis-independent representation of the electromagnetic fields and currents as used in Refs.~\cite{GolykEtAl2013,KruegerEtAl2011,KruegerEtAl2012,BimonteEtAl2017}. To make our manuscript self-consistent we will give first a brief introduction into this approach. The scattering approach starts with the Helmholtz equation~\cite{KruegerEtAl2012}
\begin{equation}
  \biggl( \mathds{H}_0  - \mathds{V} - \frac{\omega^2}{c^2} \biggr) \mathbf{E}(\mathbf{r}) = \mathbf{0}
\end{equation} 
where $\mathds{H}_0 = \nabla\times\nabla\times\mathds{1}$ and 
the potential of a scatterer with the permittivity tensor $\uuline{\epsilon}$ and permeability tensor $\uuline{\mu}$ occupying a volume $V$ is defined by
\begin{equation}
  \mathds{V} = \frac{\omega^2}{c^2} \bigl( \uuline{\epsilon}(\mathbf{r}) - \mathds{1} \bigr) + \nabla\times\biggl( \mathds{1} -\uuline{\mu}^{-1} \biggr)\nabla\times\mathds{1}.
\end{equation} 
Here we consider a homogeneous medium for the scatterer so that $ \uuline{\epsilon}(\mathbf{r}),\uuline{\mu}(\mathbf{r})$ are constant within the volume of the scatterer and equal to $\mathds{1}$ outside the scatterer so that $\mathds{V}$ is only nonzero within the scatterer.
The solution of the Helmholtz equation is determined by the Lippmann-Schwinger equation
\begin{equation}
  \mathbf{E}(\mathbf{r}) = \mathbf{E}_{0}(\mathbf{r}) + \int_V\!\!\rd r' \, \mathds{G}_0(\mathbf{r},\mathbf{r}') \mathds{V} \mathbf{E} (\mathbf{r}').
\end{equation}
Here $\mathbf{E}_{0}$ is the solution of the free Helmholtz equation and 
$ \mathds{G}_0$ is the Green tensor defined as solution of the free Helmholtz equation for a point source
\begin{equation}
   \biggl( \mathds{H}_0  - \frac{\omega^2}{c^2}\biggr) \mathds{G}_0 (\mathbf{r,r'}) = \mathds{1} \delta(\mathbf{r}-\mathbf{r}').
\end{equation}

Defining now the position basis by the ket vectors $|\mathbf{r} \rangle$ with the usual completeness
\begin{equation}
  \int \rd^3 r |\mathbf{r} \rangle \langle \mathbf{r} | = \mathds{1},
\end{equation}
and Dirac orthonormality
\begin{equation}
   \langle \mathbf{r} | \mathbf{r}' \rangle = \delta (\mathbf{r} - \mathbf{r}')
\end{equation}
and introducing the notations for vectors and tensors
\begin{align}
  \mathbf{E} (\mathbf{r}) &= \langle\mathbf{r} | \mathbf{E} \rangle, \\
  \mathds{G}(\mathbf{r},\mathbf{r}') &=  \langle \mathbf{r} | \mathds{G} | \mathbf{r}' \rangle 
\end{align}
we can bring the Helmholtz equation in a basis-independent form
\begin{equation}
   \biggl( \mathds{H}_0  - \mathds{V} - \frac{\omega^2}{c^2} \biggr) | \mathbf{E} \rangle = | \mathbf{0} \rangle
\end{equation}
where $| \mathbf{0} \rangle$ is the zero vector. The solution given by the Lippmann-Schwinger equation is then
\begin{equation}
  | \mathbf{E} \rangle = | \mathbf{E}_0 \rangle + \mathds{G}_0 \mathds{V} | \mathbf{E} \rangle.
\end{equation}
and can be expressed as
\begin{equation}
  | \mathbf{E} \rangle = | \mathbf{E}_0 \rangle + \mathds{G}_0 \mathds{T} | \mathbf{E}_0 \rangle.
\end{equation}
introducing the T-operator of the scatterer
\begin{equation}
	 \mathds{T} := \mathds{V}(\mathds{1} - \mathds{G}_0 \mathds{V})^{-1}.
\end{equation}

In the remaining part of the manuscript we will consider two compact scatterers within an arbitrary environment. In this case 
the scattering potential consists of a sum $\mathds{V}_\alpha + \mathds{V}_\beta$ of the two scatterers labeled by $\alpha$ and $\beta$. 
Note that the T-operator is not just the sum of the T-operators $\mathds{T}_{\alpha/\beta}$ of the two scatterers.
Since in the following the environment is assumed to be arbitrary we will replace $\mathds{G}_0$ and $|\mathbf{E}_0 \rangle$ be the Green 
tensor $\mathds{G}$ and the background field $|\mathbf{E}_b \rangle$ which are the solutions of the Helmholtz equations
for the considered system without the two scatterers. For example if we would consider two scatterers close to a halfspace then $\mathds{G}$ and $|\mathbf{E}_b \rangle$ would be the Green tensor and the field of the halfspace geometry, only.

\section{Total electric field}

Let's assume that we have two compact objects $\alpha$ and $\beta$ with scattering potentials $\mathds{V}_{\alpha/\beta}$, T-operators $\mathds{T}_{\alpha/\beta}$ occupying a volume $V_{\alpha/\beta}$. The two scatterers are placed in an arbitrary environment. Then the total electric field $| \mathbf{E} \rangle$ is determined by the background field $| \mathbf{E}_{\rm b} \rangle$ and the field radiated and scattered by both objects. Hence the total electrical field can be expressed by
\begin{equation}
  | \mathbf{E} \rangle = | \mathbf{E}_{\rm b} \rangle + \ri \omega \mu_0  \mathds{G} \bigl( | \mathbf{J}_\alpha\rangle + | \mathbf{J}_\beta\rangle  \bigr)
\end{equation}
where $\mathds{G}$ is the Green's tensor of the environment without the two objects $\alpha$ and $\beta$ and $\mu_0$ is the permeability of vacuum. The currents $| \mathbf{J}_{\alpha/\beta} \rangle$ are due to direct thermal radiation and the scattered radiation and can therefore be expressed by a fluctuational current and an induced current
\begin{equation}
  | \mathbf{J}_{\alpha/\beta} \rangle = | \mathbf{J}^{\rm fl}_{\alpha/\beta} \rangle + | \mathbf{J}^{\rm ind}_{\alpha/\beta} \rangle.
\end{equation}
Within the scattering approach the induced current can be expressed by the scattered field coming from the environment and the other object, i.e.\ for object $\alpha$ we have for example
\begin{equation}
   | \mathbf{J}^{\rm ind}_{\alpha} \rangle = \frac{1}{\ri  \omega\mu_0}  \mathds{T}_\alpha | \mathbf{E}_\beta \rangle
\end{equation} 
where $\mathds{T}_\alpha$ is the T-operator of object $\alpha$ describing the scattering and $| \mathbf{E}_\beta \rangle$ is the field of the environment and object $\beta$ so that
\begin{equation}
  | \mathbf{E}_\beta \rangle = | \mathbf{E}_{\rm b} \rangle + \ri \omega  \mu_0 \mathds{G} | \mathbf{J}_\beta\rangle. 
\end{equation}
This means, that at this step we ignore any self-polarisation of the object which might be included at a later stage by replacing polarisabilities by dressed polarisabilities. Now, since $ | \mathbf{J}_{\alpha} \rangle$ and $ | \mathbf{J}_{\beta} \rangle$ are related by the multiple scattering between the objects, we can express both by means of $ | \mathbf{J}_{\alpha}^{\rm fl} \rangle$,  $ | \mathbf{J}_{\beta}^{\rm fl} \rangle$, and $| \mathbf{E}_{\rm b} \rangle$.  We obtain 
\begin{equation}
	\begin{split}
		| \mathbf{J}_\alpha \rangle &= \mathds{G}^{-1} \mathds{D}_{\alpha \beta} \mathds{G} \biggl[ |  \mathbf{J}_\alpha^{\rm fl} \rangle + \mathds{T}_\alpha \mathds{G} | \mathbf{J}_\beta^{\rm fl} \rangle   \\
					    & \qquad + \frac{1}{\ri \omega \mu_0} \mathds{T}_\alpha (1 + \mathds{G} \mathds{T}_\beta) | \mathbf{E}_{\rm b} \rangle \biggr]
	\end{split}
\end{equation}
and the corresponding expression for $|\mathbf{J}_\beta \rangle$.
Hence, the total electric field can also be expressed by these quantities only and takes the form
\begin{equation}
  | \mathbf{E} \rangle = \mathds{A}_{\alpha\beta} | \mathbf{E}_{\rm b} \rangle + \ri \omega \mu_0 \biggl(\mathds{O}_\alpha \mathds{G} | \mathbf{J}^{\rm fl}_\alpha \rangle + \mathds{O}_\beta \mathds{G} | \mathbf{J}^{\rm fl}_\beta \rangle \biggr)
\end{equation}
where we have introduced the tensors
\begin{align}
 % \mathds{O}_\alpha &= \bigl(\mathds{1} + \mathds{G} \mathds{T}_\beta\bigr) \mathds{D}_{\alpha\beta}, \\
  \mathds{O}_\beta &= \bigl(\mathds{1} + \mathds{G} \mathds{T}_\alpha\bigr) \mathds{D}_{\beta\alpha}, \\
  \mathds{A}_{\alpha\beta} &= \mathds{O}_\beta \bigl(\mathds{1} + \mathds{G} \mathds{T}_\beta\bigr), 
\end{align}
and the Fabry-Perot-like ``denominator''
\begin{equation}
  \mathds{D}_{\alpha\beta} = \bigl( \mathds{1} - \mathds{G} \mathds{T}_\alpha \mathds{G} \mathds{T}_\beta \bigr)^{-1}. 
%  \mathds{D}_{\beta\alpha} &= \bigl( \mathds{1} - \mathds{G} \mathds{T}_\beta \mathds{G} \mathds{T}_\alpha \bigr)^{-1}. 
\end{equation}
The expresssions for $\mathds{O}_\alpha$ and $\mathds{D}_{\beta\alpha}$ can be obtained from $\mathds{O}_\beta$ and $\mathds{D}_{\alpha\beta}$ by interchanging $\alpha \leftrightarrow \beta$. Note, that $\mathds{D}_{\alpha\beta} \mathds{G} \mathds{T}_\alpha = \mathds{G} \mathds{T}_\alpha \mathds{D}_{\beta\alpha}$.

\section{Equilibrium correlation functions}

Since we have now the expression for the total electric field of the two objects in presence of an environment
we can determine the correlation functions of the total fields. To this end, we assume that the background field 
is in local thermal equilibrium at temperature $T_{\rm b}$ so that by using the fluctuation-dissipation theorem, 
the correlation function reads~\cite{Agarwal1975, SCHEEL}
\begin{equation}
 \llangle[\big] \mathbf{E}_{\rm b}(\mathbf{r}) \otimes \mathbf{E}_{\rm b}^\dagger(\mathbf{r}') \rrangle[\big]_{\rm leq} =  a(T_{\rm b}) \frac{\mathds{G}(\mathbf{r},\mathbf{r}') - \mathds{G}^\dagger(\mathbf{r}',\mathbf{r})}{2 \ri}
\label{Eq:correfunction}
\end{equation}
where we have introduced 
\begin{equation}
  a(T) := 2 \mu_0 \hbar \omega^2 (n(T) + 1) %\biggl( \hbar \omega n(T_b) + \frac{\hbar \omega}{2} \biggr)
\end{equation} 
with
\begin{equation}
   n(T) = \frac{1}{\re^{\hbar \omega / \kb T} - 1}.
\end{equation}
Note that we treat the fields fully quantum mechanically as operators and therefore the correlation function depends on the 
ordering. The above given anti-normally ordered correlation function is related to the normally ordered by the expression~\cite{Agarwal1975, SCHEEL}
\begin{equation}
 \llangle[\big] \mathbf{E}_{\rm b}^\dagger(\mathbf{r}) \otimes \mathbf{E}_{\rm b} (\mathbf{r}') \rrangle[\big]_{\rm leq} = \frac{n}{n + 1} \llangle[\big] \mathbf{E}_{\rm b}(\mathbf{r}) \otimes \mathbf{E}_{\rm b}^\dagger(\mathbf{r}') \rrangle[\big]_{\rm leq}^*.
\label{Eq:NormalAntinormal}
\end{equation}
Therefore, we can express all relations in terms of the anti-normally ordered correlation functions only. Eq.~(\ref{Eq:correfunction}) can be brought in the basis-independent representation and reads 
\begin{equation}
   \llangle[\big]   | \mathbf{E}_{\rm b} \rangle \langle \mathbf{E}_{\rm b} | \rrangle[\big]_{\rm leq} = a(T_{\rm b}) \frac{\mathds{G} - \mathds{G}^\dagger}{2 \ri}.
\label{Eq:FDTEb}
\end{equation}
If we further assume that the fluctuating currents within the objects are in local thermal equilibrium at temperatures $T_{\alpha/\beta}$
we obtain for the correlation functions of the fluctuating currents by using again the fluctuation-dissipation theorem
\begin{equation}
  \llangle[\big]   | \mathbf{J}^{\rm fl}_{\alpha/\beta} \rangle \langle \mathbf{J}^{\rm fl}_{\alpha/\beta} | \rrangle[\big]_{\rm leq} = \frac{1}{(\omega\mu_0)^2} a(T_{\alpha/\beta}) {\chi}_{\alpha/\beta} 
%a(T_{\alpha/\beta}) \mathds{O}_{\alpha/\beta} \mathds{G} {\chi}_{\alpha/\beta}  \mathds{G}^\dagger \mathds{O}_{\alpha/\beta}^\dagger
\label{Eq:FDTJ}
\end{equation}
where we have introduced the generalized susceptibility of particle $\alpha$ and $\beta$
\begin{equation}
    {\chi}_{\alpha/\beta} = \frac{\mathds{T}_{\alpha/\beta} - \mathds{T}_{\alpha/\beta}^\dagger}{2 \ri} - \mathds{T}_{\alpha/\beta} \frac{\mathds{G} - \mathds{G}^\dagger}{2 \ri} \mathds{T}_{\alpha/\beta}^\dagger.
\end{equation}
Since $| \mathbf{E}_{\rm b} \rangle$, $| \mathbf{J}_\alpha^{\rm fl} \rangle$, and $| \mathbf{J}_\beta^{\rm fl} \rangle$ are uncorrelated, the correlation function
of the total field $| \mathbf{E} \rangle$ is just the sum of the correlation functions of these three quantities. It is easy to show that
in global thermal equilibrium when $T_{\rm b} = T_\alpha = T_\beta \equiv T$, the sum of the three correlation functions gives for the correlation function of the total field
\begin{equation}
  \llangle[\bigl]   | \mathbf{E} \rangle \langle \mathbf{E} | \rrangle[\big]_{\rm eq} = a(T) \frac{\mathds{G}_{\rm full} - \mathds{G}^\dagger_{\rm full}}{2 \ri}
\end{equation}
where 
\begin{equation}
  \mathds{G}_{\rm full} = \mathds{A}_{\alpha\beta} \mathds{G}
\end{equation}
is the full Green's function of the two particles placed in the environment described by $\mathds{G}$. That means our local equilibrium ansatz 
in the sense of Rytov's fluctuational electrodynamics reproduces the correct global equilibrium result as it should be. Hence, as already
pointed out in Refs.~\cite{BimonteEtAl2017} the scattering approach naturally reproduces Rytov's theory, when the correct correlation functions 
for the isolated objects $\alpha$ and $\beta$ are identified. 

\section{Radiative heat transfer}

We can now determine the total power received or emitted by object $\alpha$, for instance, when assuming a local thermal equilibrium of the background field and the fluctuational currents within both objects. It is given by the symmetrically ordered expression for the dissipated power in object $\alpha$ 
\begin{equation}
   \llangle H^\alpha \rrangle = \sum_i \int_{V_\alpha}\!\!\!\rd^3 \mathbf{r} \, \llangle[\Big] \bigl\{ E_i (\mathbf{r},t) , J_{i,\alpha}(\mathbf{r},t) \bigr\}_{\rm S}\rrangle[\Big]
\end{equation}
where $E_i (\mathbf{r},t)$ and $J_{i,\alpha}(\mathbf{r},t)$ are the $i$-th component of the total field and current operators in the Heisenberg picture inside object $\alpha$. Here, $\{ A,B\}_{\rm S} = (AB + BA)/2$ symmetrizes the operator product and makes sure that in the classical limit where the operators commute we arrive at the well known classical expression for the dissipated power. Now, when switching to frequency space and assuming stationarity we obtain
\begin{equation}
  \llangle H^\alpha \rrangle  =  2 \Re \sum_{i} \! \int_0^\infty\!\!\frac{\rd \omega}{2 \pi} \!\!\int_{V_\alpha}\!\!\!\rd^3 \mathbf{r} \, \llangle[\Big] \bigl\{ E_i (\mathbf{r}) , J_{i,\alpha}^\dagger(\mathbf{r}) \bigr\}_{\rm S}\rrangle[\Big]
\end{equation}
where $E_i (\mathbf{r})$ and $J_{i,\alpha}(\mathbf{r})$ are now the $i$-th component of the Fourier components of the total field and current operators. As before we suppress the frequency argument for convenience. The brackets $\llangle[\big] \ldots \rrangle[\big]$ denoting the non-equilibrium average are to be taken as ensemble averages for the background field and the fluctuational currents being separately in local thermal equilibrium. Using relation (\ref{Eq:NormalAntinormal}) we can reexpress the above formula in the basis-independent notation as
\begin{equation}
%\begin{split}
  \llangle H^\alpha \rrangle %&= 2 \Re \int_0^\infty\!\!\frac{\rd \omega}{2 \pi} \int_{V_\alpha}\!\!\rd^3 \mathbf{r} \, \langle \mathbf{r}| \llangle[\big] | \mathbf{E}\rangle \cdot  \langle \mathbf{J}_\alpha | \rrangle[\big] | \mathbf{r} \rangle \\
                           =  \Re \int_0^\infty\!\!\frac{\rd \omega}{2 \pi} {\rm Tr}\bigl[ \llangle[\big] | \mathbf{E}\rangle \langle \mathbf{J}_\alpha | \rrangle[\big] \bigr] \frac{2 n + 1}{n + 1}.
%\end{split}
			   \label{Eq:Hkompakt}
\end{equation}
Here the factor $(2n + 1)/(n + 1)$ is in some sense symbolic, because the correct temperature in the function $n$ can only by inserted later when the non-equilibrium average is replaced by the local equilibrium averages. Since there is no correlation between the background field and the fluctuational currents this does not introduce any ambiguity, but allows us to start with a compact expression for the dissipated power in object $\alpha$ expressed by the operator trace. To this end, we have extended the integral over whole space which is possible because $\langle \mathbf{r} | \mathbf{J}_\alpha \rangle$ vanishes for $\mathbf{r} \notin V_\alpha$. This allows us then to introduced the operator trace 
\begin{equation}
   {\rm Tr}\bigl[ \mathds{A} \bigr] = \sum_i \int\!\!\rd^3 r\, \langle \mathbf{r} | \mathds{A}_{ii} | \mathbf{r} \rangle
\end{equation}
for an operator $\mathds{A}$ as defined in Ref.~\cite{KruegerEtAl2012}. Inserting the expressions for the total field $| \mathbf{E} \rangle$ and the currents $| \mathbf{J}_\alpha \rangle$ and using the fact that the background field and the fluctuating currents are uncorrelated together with the above derived correlation functions in Eqs.~(\ref{Eq:FDTEb}) and (\ref{Eq:FDTJ}) we obtain
	\begin{equation*}
		\begin{split}
			\llangle &	| \mathbf{E}\rangle \langle \mathbf{J}_\alpha |\rrangle \\ &=  
			  \ri \omega\mu_0 \mathds{O}_\alpha \mathds{G} \llangle | \mathbf{J}_\alpha^{\rm fl} \rangle \langle \mathbf{J}_\alpha^{\rm fl} | \rrangle_{\rm leq} \mathds{G}^\dagger \mathds{D}_{\alpha\beta}^\dagger {\mathds{G}^{-1}}^\dagger \\
			&\quad+ \!\!\ri \omega\mu_0\mathds{O}_\beta \mathds{G} \llangle | \mathbf{J}_\beta^{\rm fl} \rangle \langle \mathbf{J}_\beta^{\rm fl} | \rrangle_{\rm leq} \mathds{G}^\dagger \mathds{T}_\alpha^\dagger \mathds{G}^\dagger \mathds{D}_{\alpha\beta}^\dagger {\mathds{G}^{-1}}^\dagger \\
			&\quad + \!\!\mathds{A}_{\alpha\beta} \frac{\llangle | \mathbf{E}_{\rm b} \rangle \langle \mathbf{E}_{\rm b} | \rrangle_{\rm leq}}{- \ri \omega \mu_0 } (1 + \mathds{T}_\beta^\dagger \mathds{G}^\dagger) \mathds{T}_\alpha^\dagger \mathds{G}^\dagger \mathds{D}_{\alpha\beta}^\dagger {\mathds{G}^{-1}}^\dagger \\
			&= \frac{\ri}{\omega \mu_0} \biggl[ a(T_\alpha) \mathds{O}_\alpha \mathds{G} \chi_\alpha  \mathds{G}^\dagger \mathds{D}_{\alpha\beta}^\dagger {\mathds{G}^{-1}}^\dagger \\
			&\quad+ \!\!a(T_\beta) \mathds{O}_\beta \mathds{G} \chi_\beta \mathds{G}^\dagger \mathds{T}_\alpha^\dagger \mathds{G}^\dagger \mathds{D}_{\alpha\beta}^\dagger {\mathds{G}^{-1}}^\dagger \\
			&\quad + \!\!a(T_b)\mathds{A}_{\alpha\beta} \frac{\mathds{G} - \mathds{G}^\dagger}{2 \ri} (1 + \mathds{T}_\beta^\dagger \mathds{G}^\dagger) \mathds{T}_\alpha^\dagger \mathds{G}^\dagger \mathds{D}_{\alpha\beta}^\dagger {\mathds{G}^{-1}}^\dagger \biggr].
		\end{split}
	\end{equation*}
Inserting this expression into Eq.~(\ref{Eq:Hkompakt}) and keeping in mind that the factor $(2n + 1)/(n + 1)$ replaces in $a(T)$ just $n + 1$ by $2n + 1$ at the corresponding local temperature we finally arrive at
\begin{equation}
\begin{split}
   \llangle H^\alpha \rrangle &= -3 \int_0^\infty\!\!\frac{\rd \omega}{2 \pi}  \, \hbar \omega \bigl[ (n(T_\alpha) - n(T_b)) \mathcal{T}_{1}^\alpha \\
          &\,\qquad\qquad\qquad  +  (n(T_\beta) - n(T_b)) \mathcal{T}_{2}^\alpha \bigr]
\end{split}
\label{Eq:Halpha}
\end{equation}
where the transmission factors $\mathcal{T}_{1}^\alpha$ and $\mathcal{T}_{2}^\alpha$ describe the heat exchange between object $\alpha$ and the background and object $\beta$. They are defined as
\begin{align}
   \mathcal{T}_{1}^\alpha &= \frac{4}{3}\Im {\rm Tr}\bigl[\mathds{O}_\alpha \mathds{G} \chi_{\alpha} \mathds{G}^\dagger \mathds{D}_{\alpha \beta}^\dagger {\mathds{G}^{\dagger}}^{-1}  \bigr], \\
   \mathcal{T}_{2}^\alpha &= \frac{4}{3}\Im {\rm Tr}\bigl[\mathds{O}_\beta  \mathds{G} \chi_\beta  \mathds{G}^\dagger \mathds{D}_{\beta\alpha}^\dagger \mathds{T}_{\alpha}^\dagger \bigr]. 
\end{align}
The corresponding expression for $\langle H^\beta \rangle$ can be obtained by interchanging $\alpha \leftrightarrow \beta$.
Note that by convention the emitted power of object $\alpha$ is negativ, whereas the received power is positiv.
We have verified that our results are in agreement with the corresponding expressions in Refs.~\cite{GolykEtAl2013,KruegerEtAl2012}.
Obviously, in global thermal equilibrium the total power emitted or radiated by object $\alpha$ or $\beta$ vanishes as 
expected for equilibrium.  When object $\beta$ has the same temperature as object $\alpha$ then both transmission factors contribute equally.
Since in this case there is no net heat flux between the objects $\alpha$ and $\beta$, but only from object $\alpha$ to the environment, 
we can define $\mathcal{T}_{\alpha \rightarrow b} = \mathcal{T}_1^\alpha +\mathcal{T}_2^\alpha$ as the transmission factor describing the heat flux 
from $\alpha$ to the background. On the other hand, when object $\alpha$ has the same temperature as the background then the 
contribution from $\mathcal{T}_1^\alpha$ vanishes. In this situation, there is only a net heat flux 
between objects $\alpha$ and $\beta$ so that we can identify  $\mathcal{T}_{\beta \rightarrow \alpha} = -\mathcal{T}_2^\alpha$ as 
the transmission factor describing the heat flux between both objects.

\section{Green-Kubo relation}

The derivation of the Green-Kubo relation hinges now on the evaluation of quantities like
\begin{equation}
\begin{split}
  H_{\alpha\beta} &:= \int_{-\infty}^{+\infty} \!\!\rd t\, \llangle[\big] H^\alpha(t) H^\beta(0) \rrangle[\big]_{\rm eq} \\
                  &= \sum_{i,j} \int_{-\infty}^{+\infty} \!\!\!\!\rd t\, \int_{V_\alpha}\!\!\!\!\rd^3 \mathbf{r} \, \int_{V_\beta}\!\!\!\rd^3 \mathbf{r}' \,\\
                  & \qquad \llangle[\Big] \bigl\{ E_i (\mathbf{r},t) , J_{i,\alpha}(\mathbf{r},t) \bigr\}_{\rm S} \\
                  & \qquad\qquad  \times\bigl\{ E_j (\mathbf{r}',0) , J_{j,\beta}(\mathbf{r}',0) \bigr\}_{\rm S} \rrangle[\Big]_{\rm eq}
\end{split}
\end{equation}
for all combinations of $\alpha$ and $\beta$ for global thermal equilibrium at temperature $T$. 
To evaluate such expressions we have to evaluate fourth-order correlation 
functions of the total fields and the total currents. By making the assumption that the thermal radiation field has
the Gauss property~\cite{Goodman}, these correlation functions can be expressed in terms of the second-order correlation
function by virtue of the moment theorem~\cite{Goodman}. Then by switching to frequency space we can express $H_{\alpha \beta}$ by~\cite{GolykEtAl2013}
\begin{equation}
\begin{split}
  H_{\alpha\beta} &= 2 \Re \sum_{i,j} \int_0^\infty \!\!\frac{\rd \omega}{2 \pi} \, \int_{V_\alpha}\!\!\rd^3 r \int_{V_\beta}\!\!\rd^3 r'\, \frac{n}{n + 1}\\
                  &\qquad \quad\bigl[ \llangle[\big] {E}_i(\mathbf{r}) {E}^\dagger_j(\mathbf{r}') \rrangle[\big]_{\rm eq} \llangle[\big] {J}_j(\mathbf{r}') {J}_i^\dagger(\mathbf{r}) \rrangle[\big]_{\rm eq} \\
                  & \qquad\qquad +  \llangle[\big] {E}_i(\mathbf{r}) {J}^\dagger_j(\mathbf{r}') \rrangle[\big]_{\rm eq} \llangle[\big] {E}_j(\mathbf{r}') {J}_i^\dagger(\mathbf{r}) \rrangle[\big]_{\rm eq} \bigr].
\end{split}
\end{equation}
As before, we introduce the basis-independent notation
\begin{equation}
\begin{split}
    H_{\alpha\beta} &= 2 \Re \int_0^\infty \!\!\frac{\rd \omega}{2 \pi} \, {\rm Tr} \biggl[ \llangle[\big] | \mathbf{E} \rangle \langle \mathbf{E} | \rrangle[\big]_{\rm eq}   \llangle[\big] | \mathbf{J} \rangle \langle \mathbf{J} | \rrangle[\big]_{\rm eq} \\
                    &\qquad + \bigl[ \llangle[\big] | \mathbf{E} \rangle \langle \mathbf{J} | \rrangle[\big]_{\rm eq}   \llangle[\big] | \mathbf{E} \rangle \langle \mathbf{J} | \rrangle[\big]_{\rm eq} \biggr] \frac{n}{n + 1}
\end{split}
\end{equation}
by extending the volume integrals to all space and using the operator trace. In this expression the factor $n / (n + 1)$ is well defined and evaluated at the global equilibrium temperature $T$. Then by inserting the expressions for the total fields and total currents where we can use the expressions for $|\mathbf{J}_{\alpha/\beta} \rangle$ only, depending on the volume $V_{\alpha/\beta}$ considered in the integration. With the equilibrium expressions for the correlation functions in  Eqs.~(\ref{Eq:FDTEb}) and (\ref{Eq:FDTJ}) and again the assumption that the background field and the fluctuating currents are uncorrelated we arrive after a lengthy and tedious calculation at
\begin{equation}
  H_{\alpha\beta} = 3 \int_0^\infty \!\!\!\frac{\rd \omega}{2 \pi} \, (\hbar \omega)^2 n(T) \bigl[  n(T) + 1 \bigr] \bigl( \mathcal{T}_2^\alpha +  \mathcal{T}_2^\beta \bigr).
\label{Eq:Halphabeta}
\end{equation}
By comparing this result with the expression for the power received by object $\alpha$ in Eq.~(\ref{Eq:Halpha}) we obtain the 
Green-Kubo relation
\begin{equation}
  - \biggl[ \frac{\rd \llangle H^{\alpha} \rrangle}{\rd T_\beta} +  \frac{\rd \llangle H^{\beta} \rrangle}{\rd T_\alpha} \biggr]_{T_\alpha = T_\beta = T} = \frac{1}{\kb T^2} H_{\alpha\beta}. %:= \int_{-\infty}^{+\infty} \!\!\rd t\, \bigl\langle H^\alpha(t) H^\beta(0) \bigr\rangle_0
\end{equation}
This relation is not the same as in Eq.~(\ref{Eq:GreenKuboGolyk}) derived in Ref.~\cite{GolykEtAl2013}. When defining the transport coefficents as in Ref.~\cite{GolykEtAl2013} by 
\begin{equation}
   k^{\alpha}_\beta =  - \frac{\rd \llangle H^{\alpha} \rrangle}{\rd T_\beta}\biggr|_T
\end{equation}
so that  $k^{\alpha}_\beta \Delta T$ is the power emitted by object $\beta$ towards object $\alpha$ when heating it by $\Delta T$ with respect to the environment temperature $T$. With that definition we can express the Green-Kubo relation for thermal radiation as
\begin{equation}
\begin{split}
  k^\alpha_\beta + k^\beta_\alpha &=  \frac{1}{\kb T^2} H_{\alpha\beta} \\
                                  &=  \frac{1}{\kb T^2} \int_{-\infty}^{+\infty} \!\!\rd t\, \llangle[\big] H^\alpha(t) H^\beta(0) \rrangle[\big]_{\rm eq}.
\end{split}
\label{Eq:GreenKuboNeu}
\end{equation}
Therefore, the equilibrium fluctuations quantified by $H_{\alpha\beta}$ are related to both transport coefficients $k^\alpha_\beta$ and $k^\beta_\alpha$. It is evident from the above expression that $H_{\alpha\beta} = H_{\beta\alpha}$. Finally, we further find that
\begin{equation}
\begin{split}
  k^\alpha_\alpha &=  \frac{1}{2 \kb T^2} H_{\alpha\alpha} \\
                 &=  \frac{1}{2 \kb T^2} \int_{-\infty}^{+\infty} \!\!\rd t\, \llangle[\big] H^\alpha(t) H^\alpha(0) \rrangle[\big]_{\rm eq}.
\end{split}
\label{Eq:GreenKuboNN}
\end{equation}
The corresponding expression for $k^\beta_\beta$ can be obtained by replacing $\alpha \leftrightarrow \beta$. To summarize, we can regard Eq.~(\ref{Eq:GreenKuboNeu}) as the general result for all combinations of $\alpha$ and $\beta$. Just by replacing the sought for combination of indicies in that Eq.~(\ref{Eq:GreenKuboNeu}) we find the corresponding result. For example, when replacing $\beta$ in Gl.~(\ref{Eq:GreenKuboNeu}) by $\alpha$ we obtain (\ref{Eq:GreenKuboNN}), etc.

Now, by comparing our result in  Eq.~(\ref{Eq:GreenKuboNeu}) with that in Eq.~(\ref{Eq:GreenKuboGolyk}) we see that in general the relation in Eq.~(\ref{Eq:GreenKuboGolyk}) contains both transport coefficients $k^\alpha_\beta$ and $k_\alpha^\beta$ in the Green-Kubo expression when considering $H_{\alpha\beta}$ or  $H_{\beta\alpha}$. Strictly, speaking in this case our relation coincides with that of Eq.~(\ref{Eq:GreenKuboGolyk}) only if $\mathcal{T}_2^\alpha = \mathcal{T}_2^\beta$. To understand under which conditions this equality holds we first write $\mathcal{T}_2^{\alpha/\beta}$ in the following form
\begin{align}
	\mathcal{T}_{2}^\alpha &= - \frac{4}{3} {\rm Tr}\bigl[ \mathds{D}_{\beta\alpha} \mathds{G} \chi_\beta  \mathds{G}^\dagger \mathds{D}_{\beta\alpha}^\dagger \tilde{\chi}_\alpha \bigr], \label{Eq:T2alpha} \\
   \mathcal{T}_{2}^\beta &= - \frac{4}{3} {\rm Tr}\bigl[ \mathds{D}_{\alpha\beta} \mathds{G} \chi_\alpha  \mathds{G}^\dagger \mathds{D}_{\alpha\beta}^\dagger \tilde{\chi}_\beta \bigr], \label{Eq:T2beta}
\end{align}
 where we have introduced the quantity
\begin{equation}
   {\tilde{\chi}}_{\alpha/\beta} = \frac{\mathds{T}_{\alpha/\beta} - \mathds{T}_{\alpha/\beta}^\dagger}{2 \ri} - \mathds{T}_{\alpha/\beta}^\dagger \frac{\mathds{G} - \mathds{G}^\dagger}{2 \ri} \mathds{T}_{\alpha/\beta}.
\end{equation}
%Therefore, strictly speaking $\mathcal{T}_2^\alpha = \mathcal{T}_2^\beta$ can only be the same if $\mathds{T}_\alpha = \mathds{T}_\beta$, i.e. if both objects have the same geometry and material properties. But even in this case we do not necessarily have $\mathcal{T}_2^\alpha = \mathcal{T}_2^\beta$. It can only be fulfilled if the Green's tensor $\mathds{G}$ connecting object $\alpha$ and $\beta$ is the same as that one connecting $\beta$ and $\alpha$ and this is only true if the environment is reciprocal. 
{We will now show that $\mathcal{T}_2^\alpha = \mathcal{T}_2^\beta$ if $\mathds{T}_{\alpha/\beta}^t = \mathds{T}_{\alpha/\beta}$ and $\mathds{G}^t(\mathbf{r}_\alpha,\mathbf{r}_\beta) = \mathds{G}(\mathbf{r}_\beta,\mathbf{r}_\alpha)$. To this end, we start with Eq.~(\ref{Eq:T2alpha}) using the fact that the trace is invariant under transposition and the cyclic permutation property. We obtain
\begin{equation}
	\begin{split}	
   \mathcal{T}_{2}^\alpha %&= - \frac{4}{3} {\rm Tr}\bigl[ \mathds{D}_{\beta\alpha} \mathds{G} \chi_\beta  \mathds{G}^\dagger \mathds{D}_{\beta\alpha}^\dagger \tilde{\chi}_\alpha \bigr] \\
			  &= - \frac{4}{3} {\rm Tr}\bigl[ (\mathds{D}_{\beta\alpha} \mathds{G} \chi_\beta  \mathds{G}^\dagger \mathds{D}_{\beta\alpha}^\dagger \tilde{\chi}_\alpha)^t \bigr] \\
                          &= - \frac{4}{3} {\rm Tr}\bigl[ \tilde{\chi}_\alpha^t  \mathds{D}_{\beta\alpha}^* \mathds{G}^*\chi_\beta^t   \mathds{G}^t \mathds{D}_{\beta\alpha}^t   \bigr] \\
                          &= - \frac{4}{3} {\rm Tr}\bigl[  \mathds{G}^t \mathds{D}_{\beta\alpha}^t \tilde{\chi}_\alpha^t  \mathds{D}_{\beta\alpha}^* \mathds{G}^*\chi_\beta^t     \bigr].
	\end{split}
\end{equation}
At this step we assume that both objects and the environment are reciprocal. Then we have $\mathds{T}_{\alpha/\beta}^t = \mathds{T}_{\alpha/\beta}$ and $\mathds{G}^t = \mathds{G}$.  It follows that in this case $\chi_{\alpha/\beta}^t = \tilde{\chi}_{\alpha/\beta}$. It remains to show that $\mathds{D}_{\beta\alpha} \mathds{G} = \mathds{G}^t \mathds{D}_{\alpha\beta}^t$ to have $\mathcal{T}_2^\alpha = \mathcal{T}_2^\beta$. This can be done by writing this equation $\mathds{D}_{\beta\alpha} \mathds{G} = \mathds{G}^t \mathds{D}_{\alpha\beta}^t$ as $\mathds{G} {\mathds{D}_{\alpha\beta}^{-1}}^{t} = \mathds{D}_{\beta\alpha}^{-1} \mathds{G}^t$ which is simply
\begin{equation}
	\mathds{G} (\mathds{1} -  \mathds{T}_\beta^t \mathds{G}^t \mathds{T}_\alpha^t \mathds{G}^t) = (\mathds{1} - \mathds{G} \mathds{T}_\beta \mathds{G} \mathds{T}_\alpha )\mathds{G}^t. 
\end{equation}
Obviously both sides of this equation are the same if $\mathds{T}_{\alpha/\beta}^t = \mathds{T}_{\alpha/\beta}$ and $\mathds{G}^t = \mathds{G}$ so that under this condition $\mathcal{T}_2^\alpha = \mathcal{T}_2^\beta$.}
Hence, only if the objects and the environment are reciprocal our generalized Kubo-Formula retrieves the expression in Eq.~(\ref{Eq:GreenKuboGolyk}). That means that in general if only one part of the system is non-reciprocal then $\mathcal{T}_2^\alpha \neq \mathcal{T}_2^\beta$ and then the generalized Green-Kubo relation applies.

{
	Let us consider the following simple example to show that when certain approximations apply then only the environment is responsible for reciprocity or non-reciprocity of the heat transport. To this end, we consider two identical objects with $\mathds{T}_\alpha = \mathds{T}_\beta$ without making any assumption on the reciprocity of the objects. Then in general the generalized Green-Kubo relation applies, but in this specific case it can happen that the non-reciprocity is driven by the environment only. To see this we simplify first the expressions for $\mathcal{T}_2^\alpha$ and $\mathcal{T}_2^\beta$ in Eqs.~(\ref{Eq:T2alpha}) and (\ref{Eq:T2beta}) by using the single scattering approximation ($\mathds{D}_{\alpha\beta} \approx \mathds{D}_{\beta\alpha} \approx 1$) yielding
\begin{align}
	\mathcal{T}_{2}^\alpha &= - \frac{4}{3} {\rm Tr}\bigl[ \mathds{G} \chi_\beta  \mathds{G}^\dagger \tilde{\chi}_\alpha \bigr], \label{Eq:T2alpha2} \\
   \mathcal{T}_{2}^\beta &= - \frac{4}{3} {\rm Tr}\bigl[ \mathds{G} \chi_\alpha  \mathds{G}^\dagger \tilde{\chi}_\beta \bigr]. \label{Eq:T2beta2}
\end{align}
It seems that both are the same when $\chi_\alpha = \chi_\beta$ which is approximately true if the second terms in $\chi_{\alpha/\beta}$ of order $(\mathds{T})^2$ can be neglected or if both objects ``see'' the same environment. This would mean that the transport coefficients $k_\alpha^\beta$ and $k_\beta^\alpha$ are the same as well and therefore the reciprocal Green-Kubo relation would apply. But one has to keep in mind that the Green's functions in the expression for $\mathcal{T}_2^\alpha$ are those corresponding to $\mathds{G}(\mathbf{r}_\alpha,\mathbf{r}_\beta)$ and in $\mathcal{T}_2^\beta$ we have the Green's function  $\mathds{G}(\mathbf{r}_\beta,\mathbf{r}_\alpha)$. Hence, only if $\mathds{G}(\mathbf{r}_\alpha,\mathbf{r}_\beta) = \mathds{G}(\mathbf{r}_\beta,\mathbf{r}_\alpha)^t$ or $\mathds{G}(\mathbf{r}_\alpha,\mathbf{r}_\beta) = \mathds{G}(\mathbf{r}_\beta,\mathbf{r}_\alpha)$ the expressions for $\mathcal{T}_2^\alpha$ and $\mathcal{T}_2^\beta$ are the same. Therefore if the environment is reciprocal then we have $k_\alpha^\beta  = k_\beta^\alpha$ no matter if the objects themselves are reciprocal or not. But if the environment is non-reciprocal the transport coefficients are not the same. Hence for identical objects the non-reciprocity in the transport coefficients is induced by the non-reciprocal environment only if the above approximations apply. Indeed it} has recently been shown that when placing two nanoparticles in a non-reciprocal environment the transport coefficients $k_\alpha^\beta$ and $k_\beta^\alpha$ can differ a lot allowing for an efficient heat flux rectification~\cite{ZhuEtAl2018,OttetAl2019}.% \textcolor{red}{In fact, in Ref.~\cite{OttetAl2019} the above approximations were applied and it has been shown that the non-reciprocity is driven by the non-reciprocal surface waves of an nearby interface which plays the role of the environment.}

\section{Summary and Conclusions}

In summary, we have rederived the general form of the Green-Kubo relation establishing a link between the radiative heat transfer between two arbitrary compact objects in an arbitrary environment to the equilibrium fluctuations in this system. By this we have generalized a previously derived Green-Kubo relation. In this generalized relation both transport coefficients describing the heat flux from object $\alpha$ to object $\beta$ and from $\beta$ to $\alpha$ are needed. This is particularly important when heat fluxes between non-reciprocal objects or in non-reciprocal environments are considered like in magneto-optical many-body systems showing persistent heat currents and angular momenta and spins of thermal radiation~\cite{Zhu2016,ZhuEtAl2018,Silveirinha,OttSpin}, giant-magneto-resistance~\cite{Ivan2018giant,Ekeroth}, Hall effect for thermal photons~\cite{PBAHall,Ottreview} as well as heat-flux rectification by non-reciprocal surface waves~\cite{OttetAl2019}. As shown in Refs.~\cite{Zhu2016,ZhuEtAl2018,Silveirinha,OttSpin} the non-reciprocity of the considered systems results in a directional global equilibrium heat flux which can result in a directional heat flux in a non-equilibrium situation~\cite{OttSpin,PBAHall,Ottreview,OttetAl2019}. This is reflected by the generalized Green-Kubo relation where the directionality or non-reciprocity found in the transport coefficients $k_\alpha^\beta$ and $k_\beta^\alpha$ is already encoded in the equilibrium fluctuations $H_{\alpha\beta}$.

\acknowledgments
S.-A.\ B. gratefully acknowledges helpful discussions on the validity of the generalized Green-Kubo relation with Matthias Kr\"{u}ger and support from Heisenberg Programme of the Deutsche Forschungsgemeinschaft (DFG, German Research Foundation) under the project No. 404073166.

%\section{Appendix A: Retrieval of reciprocal case}
%We show that $\mathcal{T}_{2}^\alpha = \mathcal{T}_{2}^\beta$ if the environment and both objects are reciprocal, i.e.\ if $\mathbf{G}^t(\mathbf{r}_\alpha,\mathbf{r}_\beta) = \mathbf{G}(\mathbf{r}_\beta,\mathbf{r}_\alpha)$ and $\mathds{T}_{\alpha/\beta}^t = \mathds{T}_{\alpha/\beta}$. 

\end{document}